# Ultrafast carrier dynamics in pristine and FeCl$_3$-intercalated bilayer graphene

**Xingquan Zou, Da Zhan, Xiaofeng Fan, Dongwook Lee, Saritha K. Nair, Li Sun, Zhenhua Ni, Zhiqiang Luo, Lei Liu, Ting Yu, Zexiang Shen, and Elbert E. M. Chia** [a]

*Division of Physics and Applied Physics, School of Physical and Mathematical Sciences, Nanyang Technological University, 637371 Singapore, Singapore*

Ultrafast carrier dynamics of pristine bilayer graphene (BLG) and bilayer graphene intercalated with FeCl$_3$ (FeCl$_3$-G), were studied using time-resolved transient differential reflection ($\Delta R/R$). Compared to BLG, the FeCl$_3$-G data showed an opposite sign of $\Delta R/R$, a slower rise time, and a single (instead of double) exponential relaxation. We attribute these differences in dynamics to the down-shifting of the Fermi level in FeCl$_3$-G, as well as the formation of numerous horizontal bands arising from the *d*-orbitals of Fe. Our work shows that intercalation can dramatically change the electronic structure of graphene, and its associated carrier dynamics.

---

[a] Electronic mail: elbertchia@ntu.edu.sg

Graphene possesses unique electronic and structural properties that have attracted great interest experimentally and theoretically.[1,2] The unusual structural and electronic properties of graphene can be tailored chemically or structurally by intercalation, deposition of metal atoms or molecules on top, incorporation of nitrogen or boron in its structure[3] or using different substrates. The understanding and control of graphene properties can open doors to new frontiers in photonics and electronics.[4] Ultrafast measurements and theoretical calculations have been applied to study the optical and electrical properties of graphite and graphene in recent years.[5-10] Although many ultrafast studies have been done on graphite, graphene and multilayer graphene, ultrafast pump-probe measurement of $FeCl_3$-intercalated bilayer graphene ($FeCl_3$ layer sandwiched by two graphene layers) has not been reported yet. Chemical doping resulting from adsorption or intercalation may well become an important aspect of future graphene research. For example, an induced potential difference between surface and interior layers by adsorption or intercalation with $Br_2$ and $I_2$ vapors can open a band gap.[3] Also, graphite intercalated with $FeCl_3$ can cause the material to be hole-doped.[11] Superconductivity was discovered when graphite was intercalated with alkali metals and rare earth materials.[12] Thus, the understanding of electron dynamics in graphene intercalated with different materials will be important for future graphene-based device applications.

The samples used in our experiment are $FeCl_3$–G and BLG deposited on $SiO_2$/Si substrate. Experiment details and material characterization can be found in another paper.[13] The degenerate pump-probe setup[14] uses Ti:sapphire mode-locked laser with 80 MHz pulse repetition rate, generating 800 nm, 40 fs pulses. The differential reflectivity ($\Delta R/R$) of the BLG sample at different temperatures are presented in Fig. 1(a). The sign of $\Delta R/R$ is negative with a rising time of the order of pulse width and the recovery of $\Delta R/R$ exhibits two distinct time scales: a fast decay time ($\tau_{fast}$) and a slower one ($\tau_{slow}$). Our data are fitted using a bi-exponentially decaying function, $\Delta R/R = A_{fast} \exp(-t/\tau_{fast}) + A_{slow} \exp(-t/\tau_{slow})$. The time scales of $\tau_{fast}$ (< 160 fs) and $\tau_{slow}$ (1.5 – 3.4 ps) are consistent with other reports.[5-9] The processes of the relaxation can be explained by using optical phonon emission and optical phonon-acoustic phonon coupling. After the electrons are excited from the valence band to the conduction band by a pump pulse, the distribution of the electrons and holes are narrow in energy and peaked in particular directions of momentum space. Elastic as well as inelastic carrier-carrier scattering randomize the momenta and thermalize the carriers into a Fermi-Dirac distribution with a temperature much higher than the lattice temperature.[15] This initial thermalization[7] occurs on a timescale of 20 – 30 fs and is beyond our experimental resolution. After thermalization, coupling between hot electrons and optical phonons cools the electrons on a timescale $\tau_{fast}$ observed in Fig. 1(b). This is consistent with the timescale in graphite, when most of the photoexcited electron energy is transferred to the optical phonons during the first 200 fs.[7] The rough temperature-independence of the electron-phonon relaxation time $\tau_{fast}$ is consistent with the data seen in metals such as Ag and Au.[16] Subsequently, the optical phonons transfer their energies to acoustic phonons via optical-acoustic phonon coupling, which accounts for the slow process ($\tau_{slow} \sim$ few ps) shown in Fig. 1(c). Eventually, the photoexcited electrons reach thermal equilibrium with the lattice.

Figure 2(a) shows $\Delta R/R$ versus pump-probe time delay in the $FeCl_3$–G sample at different temperatures. The rise time of the signal in $FeCl_3$–G (0.2 – 0.8 ps) is much longer than that in BLG (~100 fs) and deceases with increasing lattice temperature, as shown in the inset of Fig. 2(a). The fast relaxation component found in the BLG sample disappears, leaving only the slower one in $FeCl_3$–G. The relaxation process can be fitted using a single-exponential decay function: $\Delta R/R = A_{slow} \exp(-t/\tau_{slow})$. Figure 2(b) shows that the relaxation time $\tau_{slow}$ decreases with increasing lattice temperature, similar to BLG. However, the values of $\tau_{slow}$ in $FeCl_3$–G are larger than in BLG. The difference in pump-probe data of BLG and $FeCl_3$–G originates from the difference in electronic structures. For comparison, the electronic band structures of BLG and $FeCl_3$–G were simulated by density functional theoretical (DFT) calculation with the Perdew-Burke-Ernzerhof form generalized gradient approximation (PBE-GGA) and plane-wave basis in Vienna *ab initio* simulation package (VASP) code.[17] In order to model the $FelCl_3$-G structure, a super cell with lattice constants $a =$ 12.12 Å and $c =$ 9.370 Å was constructed, where the layered $FeCl_3$ with 2×2 periods was taken as commensurate with the graphene with 5×5 periods, following the supposition of Dresselhaus *et al.*[18] The projector augmented wave (PAW) method[19] was used to describe the electron-ion interactions. A kinetic energy cutoff of 400 eV and $k$-points sampling with 0.05 Å$^{-1}$ separation in the Brillouin zone were used. The structure optimizations were carried out using a conjugate gradient algorithm with a force convergence criterion of 0.01 eV/Å. Figure 3 shows the calculated electronic band structure of FeCl$_3$–G (black line) and BLG (blue line). Note the presence of numerous horizontal bands in the FeCl$_3$–G sample which originate from the $d$ orbitals of iron. The Dirac point of the FeCl$_3$–G is now located at ~+1 eV, demonstrating that FeCl$_3$ intercalated layer causes the material to become hole-doped.

The presence of a FeCl$_3$ layer sandwiched between two graphene layers can accept electrons from the top and bottom graphene layers, causing the graphene layers to become hole-doped, and modify the electronic transport properties in graphene. Moreover, the FeCl$_3$ intercalated layer introduces additional electronic states in the energy band structure which result in three effects in the pump probe results. First, these additional electronic states facilitate photoexcited electron transition between lower and higher energy levels with the assistance of optical phonons. Optical phonons can be emitted and *reabsorbed* by the carriers in the higher energy states, thus slowing down the energy relaxation of the photoexcited electrons, as shown in Fig. 1(c) and Fig. 2(b), that is, $\tau_{slow}(FeCl_3) > \tau_{slow}(BLG)$. Second, optical phonon emission and reabsorption causes the number of photoexcited electrons to be built up slowly, resulting in a longer rise time, and causes the disappearance of the initial fast relaxation process observed in BLG. This scenario is consistent with the case of sulfate-covered gold nanopaticles.[20] There the sulfate adsorbates act as a transient energy reservoir which results in the back and forth inelastic scattering of nonequilibrium electrons, causing a retardation of the internal thermalization time compared to bulk gold. Third, the opposite signs of the signals from the two samples is due to state-filling effects (in BLG) and induced probe absorption (in $FeCl_3$–G) dominating the absorption of the probe pulse. In the linear absorption range, probe differential reflectivity $\Delta R/R$ is proportional to pump fluence $F$, the relationship can be expressed as: $\Delta R/R \propto C \times \Delta\alpha \times F$, where $C$ is a parameter depends on the interaction of probe with the excited sample, $\Delta\alpha$ is sample absorbance change. In the $FeCl_3$–G sample, electrons below the Fermi level in the valence band are excited to the horizontal bands located at energy ~1.5 eV higher by the pump photons. These photoexcited electrons in the horizontal bands can then be excited to higher energy states by absorbing the incoming probe photons (induced probe absorption), and therefore resulting in a larger probe absorption. Compare this to BLG, where the absorption of pump photons by the valence band electrons to the graphene conduction band decreases the density of empty states in the conduction band, thus resulting in a smaller probe absorption. Our results show that, though the band structure of pure carbon graphene is dominated by the Dirac description, chemical modification (adsorption, substitution and intercalation) of graphene can lead to entirely new physics.

We also examined the influence of photoexcited carrier density on the relaxation time of both BLG and FeCl$_3$–G samples at room temperature, by varying the pump power. The amplitude, which is related to the photogenerated carrier density, increases linearly with pump power for both samples, as shown in Figs. 4(a), 4(b) and 4(c), showing that we are not saturating the optical transitions in our experiments. The $\tau_{fast}$ of BLG in Fig. 4(d) increases with pump fluence at low fluence, and is attributed to the coupling between electrons and optical phonons — the number of optical phonons created by hot electron relaxation is proportional to the pump fluence, and these optical phonons can be reabsorbed by the electrons, slowing the rate of hot electron cooling. This behavior of $\tau_{fast}$ versus pump fluence also agrees with theoretical calculation at low photoexcited electron density.[21] The behaviors of $\tau_{slow}$ in both samples are very similar, as seen in Figs. 4(e) and 4(f). The value of $\tau_{slow}$ increases at low fluence and become fluence-independent beyond a fluence of 3.8 µJ/cm$^2$. With increasing pump fluence, hot electron relaxation creates a larger number of optical phonons. These optical phonons can couple energy back into the electron distribution by being reabsorbed, thereby slowing the rate of electron cooling. This accounts for the increase in $\tau_{slow}$ with increasing pump fluence. At high pump fluences, hot electrons and optical phonon modes reach a quasi-equilibrium. Here the decay of the electron energy is controlled by the relaxation of optical phonons which occurs through coupling to low-energy acoustic phonon modes. Therefore, the decay of optical phonons serves as a bottleneck for the electron energy relaxation and the limiting value of $\tau_{slow}$ at high pump fluence reflects the relaxation time for the lifetime of optical phonon modes.[10] Hence $\tau_{slow}$ level off at high pump fluence.

In conclusion, we have investigated the ultrafast carrier dynamics in BLG and FeCl$_3$–G grown on SiO$_2$/Si substrates from 10 K to room temperature. A fast relaxation process followed by a slower one are found in the BLG sample. However, in FeCl$_3$–G sample, different carrier dynamics are observed: an opposite sign of $\Delta R/R$, a slower rise time, and a single (instead of double) exponential relaxation. We attribute these differences in dynamics to the down-shifting of the Fermi level in FeCl$_3$-G, as well as the formation of numerous horizontal bands arising from the *d*-orbitals of Fe. Our work shows that intercalation can dramatically change the electronic structure of graphene and its associated carrier dynamics.

The authors acknowledge support from Singapore Ministry of Education Academic Research Fund Tier 1 (RG41/07) and Tier 2 (ARC23/08), as well as the National Research Foundation Competitive Research Programme (NRF-CRP4-2008-04).

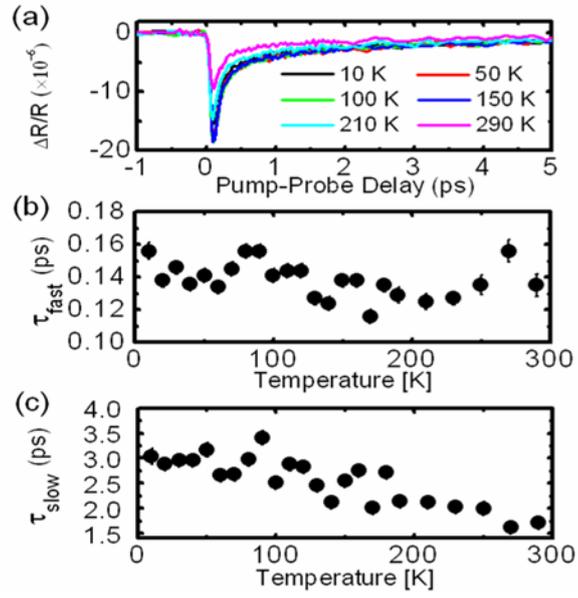

FIG. 1 (Color online) (a) Differential reflectivity $\Delta R/R$ versus pump-probe delay of BLG at different temperatures. (b) $\tau_{fast}$ and (c) $\tau_{slow}$ evolve with lattice temperatures.

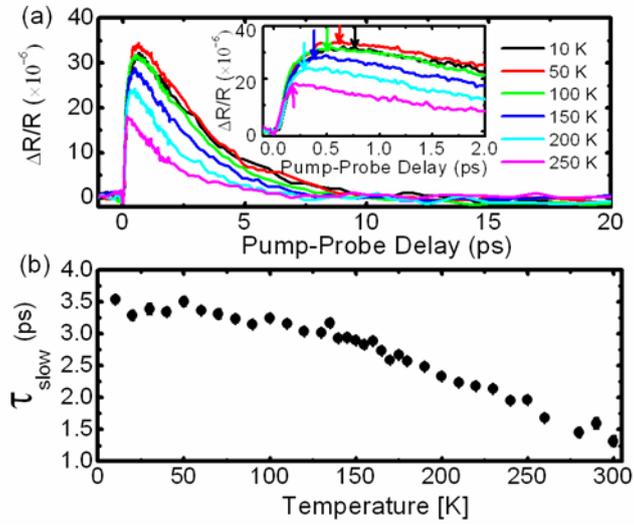

FIG. 2 (Color online) (a) Differential reflectivity $\Delta R/R$ versus pump-probe delay of FeCl$_3$–G at different temperatures. Inset: region expanded near the peak position. (b) Temperature dependence of the single relaxation time $\tau_{slow}$. Arrows show the peaks of the signals.

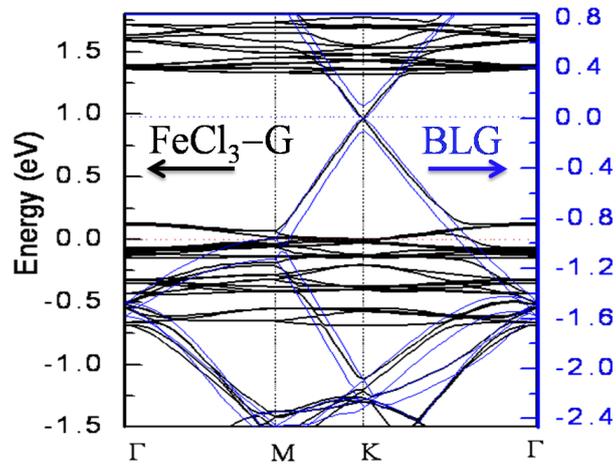

FIG. 3 (Color online) Calculated band structure of the FeCl$_3$–G (black) and BLG (blue). The horizontal bands originate from $d$ orbitals of iron and the Dirac point is located at ~+1 eV after FeCl$_3$ intercalation.

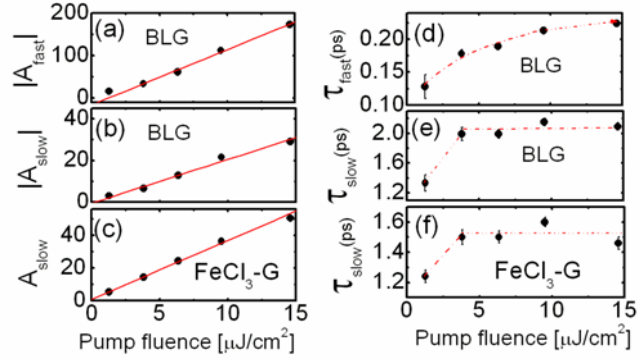

FIG. 4 (Color online) Pump fluence dependence of the amplitude (a) $|A_{fast}|$, (b) $|A_{slow}|$ in BLG and (c) $A_{slow}$ in FeCl$_3$–G. (d) The value of $\tau_{fast}$ increases with the pump fluence. The behaviors of $\tau_{slow}$ with pump fluence in BLG and FeCl$_3$–G are presented in (e), (f), respectively. The red solid lines are linear fits, and the dash dot lines guide to the eyes.